\documentclass[twoside,a4paper,11pt]{sea10}
\usepackage{graphicx}
\usepackage{hyperref}
\usepackage{movie15}
\begin{document}
\pagenumbering{arabic}
\pagestyle{myheadings}
\thispagestyle{empty}
{\flushleft\includegraphics[width=\textwidth,bb=58 650 590 680]{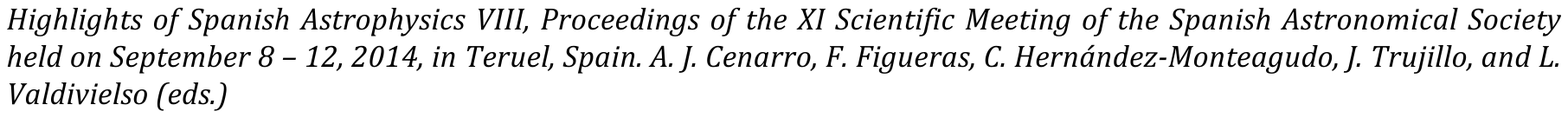}}
\vspace*{0.2cm}
\begin{flushleft}
{\bf {\LARGE
%
The IACOB spectroscopic database: recent
updates and first data release
%
}\\
\vspace*{1cm}
%
S.~Sim\'on-D\'iaz$^{1,2}$,
I.~Negueruela$^{3}$, 
J.~Ma\'iz Apell\'aniz$^{4}$
N. Castro$^{5}$,
A. Herrero$^{1,2}$,
M. Garcia$^{6}$,
J.~A.~P\'erez-Prieto$^{1,2}$,
N.~Caon$^{1,2}$,
J.~M. Alacid$^{4}$,
I. Camacho$^{1,2}$,
R. Dorda$^{3}$,
M. Godart$^{1,2}$,
C. Gonz\'alez-Fern\'andez$^{7}$,
G. Holgado$^{1,2}$, and
K. R\"ubke$^{1,2}$
%
}\\
\vspace*{0.5cm}
%
$^{1}$
Instituto de Astrof\'isica de Canarias, E-38200 La Laguna, Tenerife, Spain\\
$^{2}$
Departamento de Astrof\'isica, Univ. de La Laguna, E-38205 La Laguna, Tenerife, Spain\\
$^{3}$
Departamento de F\'isica, Ingenier\'ia de Sistemas y Teor\'ia de la Se\~nal, Escuela 
Polit\'ecnica Superior, University of Alicante, Apdo.~99, E-03080 Alicante, Spain\\
$^{4}$
Centro de Astrobiolog\'ia (CSIC-INTA), ESAC Campus, PO Box 78, E-28691 
Villanueva de la Ca\~nada, Madrid, Spain\\
$^{5}$
Argelander-Institut f\"ur Astronomie der Universit\"at Bonn, Auf dem H\"ugel 71, D-53121 Bonn, Germany\\
$^{6}$
Centro de Astrobiología, CSIC-INTA. Ctra. Torrej\'on a Ajalvir km.4, E-28850 Torrej\'on de Ardoz, Madrid, Spain\\
$^{7}$
Institute of Astronomy, Univ. of Cambridge, Madingley Road, Cambridge, CB3 0HA, UK
%
\end{flushleft}
%
\markboth{
The IACOB spectroscopic database
}{ 
%
Sim\'on-D\'iaz et al.
%
}
\thispagestyle{empty}
\vspace*{0.4cm}
\begin{minipage}[l]{0.09\textwidth}
\ 
\end{minipage}
\begin{minipage}[r]{0.9\textwidth}
\vspace{1cm}
\section*{Abstract}{\small
%
The IACOB project is an ambitious long-term project which is contributing to step forward in our knowledge about the physical properties and evolution of Galactic massive stars. The project aims at building a large database of high-resolution, multi-epoch, spectra of Galactic OB stars, and the scientific exploitation of the database using state-of-the-art models and techniques. In this proceeding, we summarize the latest updates of the IACOB spectroscopic database and highlight some of the first scientific results from the IACOB project; we also announce the first data release and the first public version of the {\tt iacob-broad} tool for the line-broadening characterization of OB-type 
spectra.
%
\normalsize}
\end{minipage}
%
%
%
\section{Introduction \label{intro}}
Massive stars have been many times claimed as Cosmic Engines and Gifts of Nature for the study of the Universe, from the Solar neighbourhood to the large-z Universe. The complete understanding of the physical properties and evolution of massive stars (and their interplay with the interstellar medium) is crucial for many fields of Astrophysics and, ultimately, to understand the evolution of the Cosmos. In spite of the remarkable advances in the modelling and spectroscopic analysis techniques of these stars in the last two decades, our knowledge of these important (but complex) astrophysical objects has been, until very recently, limited to conclusions extracted from the analysis of single-epoch medium resolution spectroscopic observations of relatively small samples. 

To step forward in this field, it has become crucial to increase the number of analysed objects in the Milky Way and other metallicity environments, also considering the information provided by multi-epoch photometric, spectroscopic (multi-wavelength) and spectropolarimetric observations. Only such a complete observational dataset allows the investigation of the impact that binarity/multiplicity, magnetic fields, and stellar oscillations (in addition to mass, metallicity, stellar winds and rotation) have on the physical properties of massive stars, as well as the evolution of these important astrophysical objects.

The IACOB project is an ambitious long-term project which is contributing to the new era of investigation of massive stars by concentrating on Galactic OB stars. In particular, the project aims at building a large database of high-resolution, multi-epoch, spectra of Galactic OB stars (the IACOB spectroscopic database), and the scientific exploitation of the database using state-of-the-art models and techniques. The main drivers of the project are [A] to increase the statistics of Galactic OB stars with accurate physical parameters and abundances, [B] to investigate in detail some of the still open questions in our knowledge of the physical properties and evolution 
of massive stars (e.g., the actual masses of these stars, the nature of the O Vz stars, the driving mechanism of stellar winds in the weak wind regime, the uncertain stage of B supergiants in an evolutionary context, the actual rotational velocities of massive stars, the physical origin of the macroturbulent broadening, and the impact of rotation/binarity/pulsations on the evolution of massive stars), and [C] to provide an unique ground-based spectroscopic database supporting/complementing the future Gaia database. During its development, the IACOB project has also established synergies with the Galactic O-Star Spectroscopic Survey (GOSSS, \cite{Mai11}), three other high-resolution 
sister surveys (OWN \cite{Bar10,Bar14}, NoMaDS \cite{Mai12, Pel12}, and CAF\'E-BEANS \cite{Neg14}), and
the VLT-FLAMES Tarantula Survey \cite{Eva11}.

The IACOB spectroscopic database was last described in detail in \cite{Sim11c}; however, the number of spectra in the database has been almost doubled since then, and its scientific drivers extended. In this contribution to the XI Scientific Meeting of the Spanish Astronomical Society, we present an updated description of the IACOB database, and also announce the first data release and the first public version of the {\tt iacob-broad} tool for the line-broadening characterization of O- and B-type stars.

\section{The IACOB spectroscopic database \label{intro}}
\subsection{Main drivers and updates since 2011}

\begin{table}[t!] 
\caption{Summary of observations in the IACOB spectroscopic database (as for Dec. 2014)} 
\center
\center
\begin{tabular}{cccc} 
\hline\hline 
           & FIES@NOT2.5m & HERMES@Mercator1.2m & \\  
           & R\,=\,25000/46000 & R\,=\,85000 & \\
           & 3600\,--\,7200\,\AA & 3750\,--\,9000\,\AA & All \\
\hline 
\# spectra & 1755 & 1950 & 3705\\ 
\# O stars (O4\,--\,O9.7) & 152 & 49 & 157 \\
\# B stars (B0\,--\,B9)   & 212 & 216 & 374\\
\hline
\end{tabular} 
\label{tab1} 
\end{table}

At present, the IACOB spectroscopic database comprises $\sim$3705 spectra of 531 bright Northern Galactic
O- and B-type stars obtained with the FIES and HERMES high-resolution spectrographs attached to the
2.5m Nordic Optical Telescope and 1.2m Mercator Telescope, respectively. A summary of the compiled observations
(as for December 2014) is presented in Table \ref{tab1}. The database has certainly surpassed its 
initial aims, while its main philosophy still remains. The initial motivation
of building a multi-epoch (at least 3 epochs), homogeneous (FIES) high-resolution database of O-type 
stars brighter than V=9 and observable from El Roque de los Muchachos Observatory\footnote{The list of targets 
was build from the \href{http://ssg.iaa.es/en/content/galactic-o-star-catalog/}{\tt Galactic O-Star Catalogue (GOSC)} \cite{Mai13}.}  was fulfilled in February 2013. In the meanwhile, and after then, the IACOB spectroscopic database has also benefited from many 
other observing campaigns on related projects which are briefly described below:

\begin{itemize}
\item {\em IACOB-sweG (IACOB supplemented with an extension to the Gaia spectral range)}: This dataset comprises $\sim$100
bright Galactic stars traditionally considered as standards for spectral classification
in the O4\,--\,A0 spectral type domain. This observational project, led by I. Negueruela, aims at building
a modern high-resolution atlas for spectral clasification of O and B stars, with especial emphasis in 
the Gaia-RVS spectral window (8450\,--\,8740\,\AA). The HERMES spectrograph was considered for this project,
since FIES spectra do no include the Gaia spectral range.
\item {\em The single snap-shot view of macroturbulent broadening in OB stars}: This dataset, which includes at least one FIES
or HERMES spectrum of a large sample of bright early to late B-type stars, was compiled to systematically extend 
the investigation of the characteristics and presence of the macroturbulent broadening to the whole O and
B star domain.
\item {\em The time-dependent view of macroturbulent broadening}: This line-broadening, which still lacks from
an observational confirmation of its physical origin, has been proposed to be a spectroscopic signature of the collective effect of pulsations in 
hot massive stars \cite{Aer09}. To investigate this hypothesis, we are obtaining FIES+HERMES time-series observations of
a small sample of selected targets showing an important macroturbulent broadening contribution. The star sample 
includes several O stars and B Supergiants. For some of the targets we already count on more than 150 spectra
gathered during the last 5 years.
\item {\em Abundances in B-type main sequence stars}: During one of the initial IACOB runs, we obtained spectra for a 
sample of 13 early B-type stars (with low projected rotational velocities) in the Orion OB1 star forming region to 
perform a chemical abundance analysis. More recently, we have continued observing other bright Northern early and mid 
B-type stars (in collaboration with M.F. Nieva) aiming at increasing the number of B-type stars in the Solar 
neighbourhood with reliable stellar parameters and abundances.
\item {\em Binary status of Galactic B Supergiants}: One of the last extensions of the IACOB database refers
to HERMES multi-epoch observations of Galactic B Supergiants, aimed at detecting and investigating binaries among the evolved descendants
of O-type stars.
\end{itemize}

\subsection{First scientific results using IACOB spectra}

The analysis of part of the spectroscopic dataset described above has already led to interesting contributions to the field of massive stars regarding topics as the actual rotational velocities in OB stars \cite{Sim14a}, the scatter in the spectral type vs. effective temperature calibrations \cite{Sim14b}, the present-day chemical composition of the Orion star forming region \cite{Sim10a, Sim11a, Nie11}, the nature of the macroturbulent broadening and its possible connection to stellar oscillations \cite{Sim10b, Sim14c}, the detection and characterization of new massive binary/multiple systems \cite{Sim11b, Sim14d} (see also I. Negueruela, these proceedings), or the use of spectroscopic observational constraints to check the reliability of modern state-of-the-art evolutionary codes \cite{Cas14}.

\subsection{Getting ready to make the IACOB deliverables available to the community}

The IACOB project now counts on a webpage\footnote{\href{http://www.iac.es/proyecto/iacob}{\tt www.iac.es/proyecto/iacob}} describing the main goals and results of the project. It also includes links to some tools of interest for the quantitative spectroscopic analysis of O and B stars (some of them developed in the framework of the IACOB project) and an interface to have access to the IACOB spectra. This interface\footnote{The IACOB archive system is based on the \href{http://svo.cab.inta-csic.es/docs/index.php?pagename=Projects/Astronomical\_Data\_Centres/Publishing}{{\tt MySpec}} tool developed by the Spanish Virtual Observatory to facilitate the publication of 1-D spectra in the VO.}, has been designed to either show the complete (updated) IACOB database or perform more detailed searches by target, spectral class, data release, instrument and/or spectral resolving power. 
The different data releases will be conveniently announced; in the meanwhile, people interested in especific (samples of) spectra can contact the PI of the project by mail (ssimon@iac.es).

\subsubsection{The IACOB spectroscopic database: First data release (DR1)}

The first data release (DR1) of the IACOB spectroscopic database is now available at {\href{http://vivaldi.ll.iac.es:8080/iacob/jsp/search.jsp}{\tt http://vivaldi.ll.iac.es:8080/iacob/jsp/search.jsp}\footnote{The interface is also accesible via the main IACOB webpage}. This release, which includes 656 spectra of 73 stars compiled 
during 8 nights in Nov. 2008 and Nov. 2009, mainly refers to the spectroscopic time-series of O and early-B Supergiants
presented in \cite{Sim10b}, and the spectra of 13 early-B dwarfs in the Orion OB1 star-forming region analyzed in \cite{Sim10a} and \cite{Nie11}. The later have been also considered for the studies performed by \cite{Irr14} and \cite{Nie14}.

\subsection{The {\tt iacob-broad} tool now available}

The IACOB project will also made available for the community some of the tools developed and used in the framework
of the project. The first one we deliver is {\tt iacob-broad}, a versatile and user-friendly IDL tool for the line-broadening characterization of OB-type spectra. The tool, described in detail in \cite{Sim14a} and based on a combined Fourier transform (FT) + goodness-of-fit (GOF) methodology, allows for the determination of the projected rotational velocity ($v$~sin$i$) and the macroturbulent broadening ($v_{\rm m}$) for a variety of situations. The first public version 
of the tool is now also available in the \href{http://www.iac.es/proyecto/iacob}{IACOB webpage}. 

%
%
\small  
%
\section*{Acknowledgments}   
%
The IACOB spectroscopic database is based on observations made with the Nordic Optical Telescope, operated by the Nordic Optical 
Telescope Scientific Association, and the Mercator Telescope, operated by the 
Flemish Community, both at the Observatorio del Roque de los Muchachos (La Palma, Spain) 
of the Instituto de Astrofisica de Canarias.
SS-D kindly acknowledges the staff at these telescopes for
their professional competence and always useful help during more than 80 observing nights between 
2008 and 2014. This work has been funded by the Spanish Ministry of Economy and
Competitiveness (MINECO) under the grants 
AYA2010-17631, 
AYA2010-15081, 
AYA2010-21697-C05-01, 
AYA2012-39364-C02-01/02, 
FIS2012-39162-C06-01 
ESP2013-47809-C3-1-R 
and 
Severo Ochoa SEV-2011-0187, 
and by the Canary Islands Government under grant 
PID2010119. This research has made use of the Spanish Virtual
Observatory (\href{http://svo.cab.inta-csic.es}{http://svo.cab.inta-csic.es}) supported from the Spanish
MINECO through grant AYA2011-24052

%

%
\end{document}